\documentclass{ws-procs9x6}

\newcommand{\refeq}[1]{(\ref{#1})}
\def\etal {{\it et al.}}

\begin{document}


\def\slashi#1{\rlap{\sl/}#1}
%
\def\slashii#1{\setbox0=\hbox{$#1$}             
   \dimen0=\wd0                                 
   \setbox1=\hbox{\sl/} \dimen1=\wd1            
   \ifdim\dimen0>\dimen1                        
      \rlap{\hbox to \dimen0{\hfil\sl/\hfil}}   
      #1                                        
   \else                                        
      \rlap{\hbox to \dimen1{\hfil$#1$\hfil}}   
      \hbox{\sl/}                               
   \fi}                                         %
%
\def\slashiii#1{\setbox0=\hbox{$#1$}#1\hskip-\wd0\hbox
to\wd0{\hss\sl/\/\hss}}
%
\def\slashiv#1{#1\llap{\sl/}}
\def\Lpo{L_+^\uparrow}

\title{A QUANTUM FIELD MODEL FOR TACHYONIC NEUTRINOS 
WITH LORENTZ SYMMETRY BREAKING}

\author{MAREK J.\ RADZIKOWSKI}

\address{Dept.\ of Physics and Astronomy, University of British Columbia\\
Vancouver, B.C.\ V6T 1Z1, Canada\\
E-mail: radzik@physics.ubc.ca}

\begin{abstract}
 A quantum field model for Dirac-like tachyons respecting a frame-dependent interpretation rule, and thus inherently breaking Lorentz invariance, is defined. It is shown how the usual paradoxa ascribed to tachyons, instability and acausality, are resolved in this model, and it is argued elsewhere that Lorentz symmetry breaking is necessary to permit perturbative renormalizability and causality. Elimination of negative-normed states results in only left-handed particles and right-handed antiparticles, suitable for describing the neutrino. In this context the neutron beta decay spectrum is calculated near the end point for large, but not ultrarelativistic preferred frame speed, assuming a vector weak interaction vertex. 
\end{abstract}

\bodymatter

\section{Introduction}\label{intro}
The purpose of this talk is very briefly to introduce a quantum field theoretic model of Dirac-like tachyons (or {\it Dirachyons}), which must necessarily incorporate spontaneous Lorentz symmetry breaking (LSB),\footnote{An analogous approach is offered by Ciborowski and Rembieli\'nski.\cite{CibRem96}} and to derive best fit curves for beta decay, which experimentalists may use to test the {\it tachyonic neutrino hypothesis}.\cite{CHK85} The metric signature is $(+,-,-,-)$ and $c=\hbar=1$. The proper, orthochronous Lorentz group is denoted by $\Lpo$. An element $\Lambda\in\Lpo$ is simply called a Lorentz transformation here, or LT for short.\footnote{The LTs herein are considered to be {\it observer} or {\it passive} transformations.}

\begin{figure}
\psfig{file=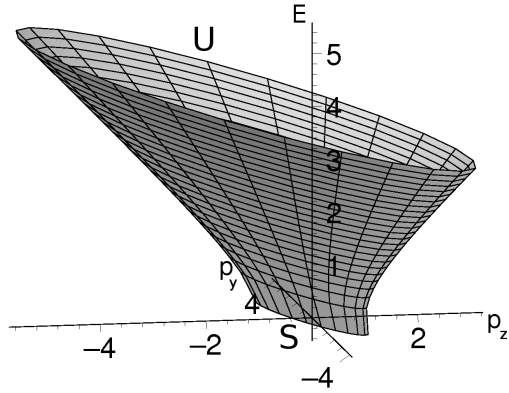,width=4.5in}
\caption{Allowed energy-momenta for one-particle states, $m=1, \beta=0.3$.}
\label{fig1}
\end{figure}

Figure\ \ref{fig1} depicts the energy-momenta of (anti-)particles allowed in this model, in a typical inertial frame ${\cal O}$. The spectrum is the upper half of the one-sheeted mass hyperboloid $E^2-{\bf p}^2 = -m^2$, sliced in two by the hyperplane  $E+\beta p_z=0$. In the {\it preferred tachyon frame} ${\cal T}={\cal O}'$, an observer views this hyperplane as $E'=0$.\footnote{Henceforth an observer in ${\cal T}$ always uses primed coordinates.} Here, ${\cal T}$'s velocity relative to ${\cal O}$ has been chosen to be $-\beta \hat{\bf z}$ in ${\cal O}$. 

Thus we have a {\it frame-dependent interpretation rule}. This is akin to the `Reinterpretation Principle,'\cite{BilDesSud62} by which a negative energy tachyon travelling backward in time, is (classically) equivalent to a positive energy (anti-) tachyon moving forward in time, with opposite momentum, etc. However, in the QFT, instead of insisting that all tachyons have positive energies in all frames, one henceforth requires this in ${\cal T}$ (mod SO(3)), and allows negative energies in all other frames, consistent with the action of $\Lpo$, without further reinterpretation, i.e., the corresponding positive energies with opposite momenta are missing from the spectrum.

\section{Definition of quantum field model}
One seeks to quantize the equation,\footnote{A similar 
prescription works for scalar tachyons obeying the Klein-Gordon equation with negative mass-squared term.} originally due to Tanaka\cite{Tan60},
\begin{equation}
 \left(i\slashi\partial - m\gamma_5\right)\psi=0\;,
\label{Dchn}
\end{equation}
which is derivable from the Hermitian Lagrangian density ${\cal L}=i\bar{\psi}\gamma_5\slashi\partial\psi - m\bar{\psi}\psi$. In ${\cal O}$, an {\it indefinite} inner product is defined for solutions $u_1(x),u_2(x)$ of Eq.\ \ref{Dchn}:
\begin{equation}
 \left(u_1,u_2\right) \equiv -\int_{t'=a'} d^3{\bf x}'\ u_1^{\prime\dagger}(x')\gamma_5 u'_2(x')\;.
 \label{IP}
\end{equation} 
Here, a $\Lambda\in\Lpo$ is chosen so that $x=\Lambda x'$, and $u'_1(x'), u'_2(x')$ are found so that $u_i(x) = D(\Lambda)u'_i(x'),\ i=1,2$, where $D$ is the $(1/2,0)\oplus(0,1/2)$ representation of $\Lpo$, i.e., $(u_1,u_2)$ is invariant under {\it observer} LTs. Adhering to the above interpretation rule, the ansatz for $\hat\psi(x)$ in ${\cal O}$ is
\begin{equation}
 \hat\psi(x) = \sum_s\int_{p\in U}d\nu(p)\left[u_{p,s}(x) a_{p,s} + v_{p,s}(x) b_{p,s}^\dagger\right]\;,
\end{equation}
where $d\nu(p)$ is the invariant measure on the mass hyperboloid. Also, $u_{p,s}(x)$ [$v_{p,s}(x)$] are `upper' [`lower'] energy solutions of Eq.\ \refeq{Dchn} of the form $\exp(-ip\cdot x)u_s(p)$, [$\exp(ip\cdot x)v_s(p)$], for $p\in U$, obtained from suitably normalized positive [negative] energy solutions of $({\slashiv p}\,{}' - m\gamma_5)u'_s(p')=0$, [$({\slashiv p}\,{}' + m\gamma_5)v'_s(p')=0$] in ${\cal T}$ via $D(\Lambda)$. The parameter $s=\pm$ denotes helicity eigenstates. The creation/annihilation operators $a_{p,s}, b_{p,s}, a_{p,s}^\dagger, b_{p,s}^\dagger$ satisfy the anticommutation relations \begin{equation}
\left\lbrace a_{p,s},a_{q,s'}^\dagger\right\rbrace=\left\lbrace b_{p,s},b_{q,s'}^\dagger\right\rbrace=\pm 2E_p\delta_{ss'}\delta^{(3)}({\bf p}-{\bf q}),
\end{equation} 
and all other anticommutators vanish. Here, $E_p\equiv\sqrt{{\bf p}^2-m^2}$, $|{\bf p}|\ge m$, and the sign is negative for the negative normed modes. Since the anticommutators are positive definite, one must eliminate such modes from the field operator, yielding
\begin{equation}
 \hat\psi(x)=\int_{p\in U}d\nu(p)\left[ u_{p,-}(x) a_{p,-} + v_{p,+}(x) b_{p,+}^\dagger\right]\;.
\end{equation}The vacuum state $\left|0\right\rangle_{\cal O}$ is defined by $a_{p,s}\left|0\right\rangle_{\cal O} = b_{p,s}\left|0\right\rangle_{\cal O} = 0$, for all $p\in U$. The subscript is included since this vacuum is not $\Lpo$ invariant. The $n$-point functions (expectation values of products of field operators which in turn define a QFT\cite{Wig56}), and Green's functions may then be computed in ${\cal O}$. With the appropriate sign in front of the classical Hamiltonian density, the second quantized Hamiltonian (defined analogously to Eq.\ \refeq{IP}) becomes
\begin{equation}
 \hat{H} = \int_{p\in U} d\nu(p)p^0\left(a_{p,-}^\dagger a_{p,-} + b_{p,+}^\dagger b_{p,+}\right),
\end{equation}
which is positive semidefinite only in ${\cal T}$. Furthermore, there remains a consistent notion of lepton number, namely the quantization of the inner product,
Eq.\ \refeq{IP},
\begin{equation}
 \hat{Q} = \int_{p\in U}d\nu(p)\left(a_{p,-}^\dagger a_{p,-} - b_{p,+}^\dagger b_{p,+}\right)\;. 
\end{equation}

\section{Resolution of paradoxa}
The usual difficulties ascribed to tachyons are instability (of the vacuum) and acausality. There are two possible kinds of instability: (1) due to exponentially growing/decaying modes in time, and (2) due to negative real energies unbounded from below, leading to lack of perturbative renormalizability. Instabilities of type (1) are excluded from this model by fiat, since they would correspond to unphysical imaginary energies.\footnote{These could not be excited perturbatively from real energies, since they would require a four-point interaction vertex.} An instability of type (2) is avoided due to the spectral cutoff. Note that requiring the usual Lorentz covariance of the two point function would lead to a {\it non-Hadamard} state having an ill-defined stress-energy tensor, while utilizing the cutoff is conjectured to permit renormalizibility when the free model is incorporated into an interacting theory.\cite{Rad08b} The cutoff also prevents the creation of causality-violating devices, such as `anti-telephones.'\cite{Rad08b} Furthermore, the usual connection of spin and statistics is maintained in this model. This differs from Feinberg,\cite{Fei67} who evidently ruled out commutation relations for the scalar case due to a sign problem that can be traced back to the use of an inappropriate surface of integration with which to evaluate inner products. In the present case, one desires a regular massless model with broken parity in the limit as $m\to 0$, which would require the usual spin-statistics connection to hold for this tachyonic model.   

\section{Results}
To evaluate the rate for neutron beta decay ($n\longrightarrow p+e+\bar\nu$), one replaces the Feynman diagram vertex factor $-i(g_W/\sqrt{2})((1-\gamma^5)/2)\gamma^\mu$ by $-i(g_W/\sqrt{2})\gamma^\mu$, since parity breaking is already accounted for in the free model. With neutron, antineutrino, proton and electron momenta denoted $p_1, p_2,p_3,p_4$ resp., one arrives at the manifestly positive expression
\begin{eqnarray}
 \left\langle\left|{\cal M}\right|^2\right\rangle &=& 2\left(\frac{g_W}{M_W}\right)^4\left[ (p_1 \cdot p_4)({\tilde p}_2 \cdot p_3)+(p_3 \cdot p_4)({\tilde p}_2 \cdot p_1)\right.\nonumber \\
&-&\left. m_n m_p ({\tilde p}_2 \cdot p_4) - m_\nu m_e  (p_1 \cdot p_3) + 2 m_n m_p m_\nu m_e \right]\;.
\end{eqnarray}
In preferred frame coordinates, ${\tilde p'}\equiv(|{\bf p'}|, (p'_0/|{\bf p}'|){\bf p}')$ is the {\it future timelike conjugate} of the spacelike $p'=(p'_0,{\bf p}')$. To find ${\tilde p}$ in any other frame ${\cal O}$, transform the coordinates of $p$ to ${\cal T}$, take the timelike conjugate, then transform back to ${\cal O}$. With $\gamma^{-1}=\sqrt{1-\beta^2}$, $\Delta \equiv m_n-m_p$, $x \equiv (\Delta - E)/m_\nu$, and $p_e \equiv \sqrt{\Delta^2 - m_e^2}$, and assuming the detected electrons are emitted in a cone of half-angle $90^\circ$ whose axis is at polar angle $\alpha$, the differential decay rate is
\begin{eqnarray}
 \frac{d\Gamma}{dE} &=& \frac{m_\nu^2}{4\pi^3}\left(\frac{g_W}{2 M_W}\right)^4 p_e\left\lbrace\theta(x-\beta\gamma)\left[ 2m_e\sqrt{x^2 + 1} +2\left(\Delta\cdot x^2 + \frac{p_e}{4\beta}\cos\alpha\right)\right.\right.\nonumber \\ &+& \left.\left.\frac{1}{\beta}\left(\Delta - \frac{p_e}{4\beta}(1+\beta^2)\cos\alpha\right)\ln\left(\frac{1+\beta}{1-\beta}\right) \right] \right.\nonumber \\ &+& \left.\theta(\beta\gamma - x)\theta(x+\beta\gamma)\frac{1}{\beta}\left[m_e(x+\beta\sqrt{x^2+1}) + \beta\Delta\cdot x^2 + \frac{p_e}{4}\cos\alpha \right.\right.\nonumber \\ &+& \left.\left. x\sqrt{x^2+1}\left(\Delta - \frac{p_e}{4\beta\gamma^2}\cos\alpha\right) - \frac{1}{\gamma}\left(\Delta - \frac{p_e}{2\beta}\cos\alpha\right)x\right.\right.\nonumber \\ &+& \left.\left.\left(\Delta - \frac{p_e}{4\beta}(1+\beta^2)\cos\alpha\right)\ln\left(\gamma(1+\beta)(x+\sqrt{x^2+1})\right)\right]\right\rbrace\;.
\end{eqnarray}
In the approximations leading to the above, the preferred frame speed $\beta$ is allowed to be large, i.e., of order $1$, but not ultrarelativistic. An appropriately modified set of fit curves may be used to test for tachyonic neutrinos and consequent Lorentz symmety breaking, e.g., at KATRIN.\cite{KAT01}
   
\section*{Acknowledgments}
The author wishes to thank A.S.\ Wightman, K.\ Fredenhagen, G.\ Heinzelmann, B.S.\ Kay, W.G.\ Unruh, and V.A.\ Kosteleck\'y for discussions, hospitality and encouragement. Partial support was provided through a DFG at the II. Institut f\"ur Theoretische Physik, Univ.\ of Hamburg, Germany, and through a teaching post-doctoral fellowship with PHAS at UBC.

\end{document}